\newcommand*{\wn}{cm$^{-1}$}

\newcommand{\xstate}{$X^{1}\Sigma_{g}^{+}$}
\newcommand{\efstate}{$EF^{1}\Sigma^{+}_{g}$}

\newcommand*{\Hm}{H$_{2}$}
\newcommand*{\Dm}{D$_{2}$}

\documentclass[aps,prl,twocolumn,superscriptaddress,10pt,floatfix]{revtex4-1}

\usepackage{threeparttable}
\usepackage{graphicx}

\begin{document}

\title{The Fundamental Vibration of Molecular Hydrogen}
\author{G.D. Dickenson}
\affiliation{Department of Physics and Astronomy, LaserLaB, VU University, de Boelelaan 1081, 1081HV, Amsterdam, The Netherlands}
\author{M.L. Niu}
\affiliation{Department of Physics and Astronomy, LaserLaB, VU University, de Boelelaan 1081, 1081HV, Amsterdam, The Netherlands}
\author{E.J. Salumbides}
\affiliation{Department of Physics and Astronomy, LaserLaB, VU University, de Boelelaan 1081, 1081HV, Amsterdam, The Netherlands}
\affiliation{Department of Physics, University of San Carlos, Cebu City 6000, The Philippines}
\author{J. Komasa}
\affiliation{Faculty of Chemistry, A. Mickiewicz University, Grunwaldzka 6, 60-780 Pozna\'n, Poland}
\author{K.S.E. Eikema}
\affiliation{Department of Physics and Astronomy, LaserLaB, VU University, de Boelelaan 1081, 1081HV, Amsterdam, The Netherlands}
\author{K. Pachucki}
\affiliation{Faculty of Physics, University of Warsaw, Ho\.za 69, 00-681 Warsaw, Poland}
\author{W. Ubachs}
\email[Corresponding Author: ]{w.m.g.ubachs@vu.nl}
\affiliation{Department of Physics and Astronomy, LaserLaB, VU University, de Boelelaan 1081, 1081HV, Amsterdam, The Netherlands}

\begin{abstract}
The fundamental ground tone vibration of H$_2$, HD, and D$_2$ is determined to an accuracy of
$2 \times 10^{-4}$ cm$^{-1}$ from Doppler-free laser spectroscopy in the collisionless environment of a molecular beam.
This rotationless vibrational splitting is derived from the combination difference between electronic excitation
from the $X^1\Sigma_g^+, v=0$ and $v=1$ levels to a common $EF^1\Sigma_g^+, v=0$ level.
Agreement within $1\sigma$ between the experimental result and a full \emph{ab initio} calculation provides a stringent test of quantum electrodynamics in a chemically-bound system.
\end{abstract}

\maketitle
Quantum electrodynamics (QED), the fully quantized and relativistic version of electromagnetism, solves
the problem of infinities associated with charged point-like particles and includes the effects of spontaneous particle-antiparticle generation from the vacuum. QED is tested to extreme precision by comparing values for the electromagnetic coupling constant $\alpha$ obtained from measurements of the $g$-factor of the electron \cite{Hanneke2008} and from interferometric atomic recoil measurements \cite{Bouchendira2011}. These experiments and the Lamb shift measurements in atomic hydrogen \cite{Parthey2011,Schwob1999} have made QED the most accurately tested theory in physics. Concerning molecules, significant progress has been made recently in theoretical \cite{Korobov2008} and experimental \cite{Koelemeij2007,Bressel2012} investigations of QED phenomena in the HD$^+$ molecular ion, where multiple angular momenta (rotational, electronic and nuclear spins) play a role. Neutral hydrogen has also recently been targeted for QED-tests, via a measurement of the dissociation energy of the H$_2$ \cite{Liu2009}, HD~\cite{Sprecher2010}, and D$_{2}$ \cite{Liu2010} molecules, and the experimental determination of rotationally excited quantum levels in H$_2$ \cite{Salumbides2011}.

The rotationless fundamental ground tone (\emph{i.e.} the vibrational energy splitting between the $v''=0, J''=0$ and $v'=1,J'=0$ quantum states) of the neutral hydrogen molecule is an ideal test system for several reasons. The total electronic angular momentum is zero for the \xstate\ ground state and the total nuclear spin for the rotationless $J=0$ state of para-\Hm\ is also zero resulting in a simple spectrum without hyperfine splitting. The hyperfine splitting is extremely small in HD (down to the Hz level~\cite{Oddershede1988}) and \Dm\ in the absence of an $\vec{I} \cdot \vec{J}$ interaction for the $J=0$ ground state.
The recent progress in theory allows for calculations involving relativistic and QED-effects up to order $\alpha^4$ \cite{Piszczatowski2009,Komasa2011}. Energy contributions in the calculation cancel to a large degree for the fundamental ground tone, leading to a significant reduction in the uncertainty, thereby allowing for accurate QED tests.

\begin{figure}[t]
\includegraphics[width=1\linewidth]{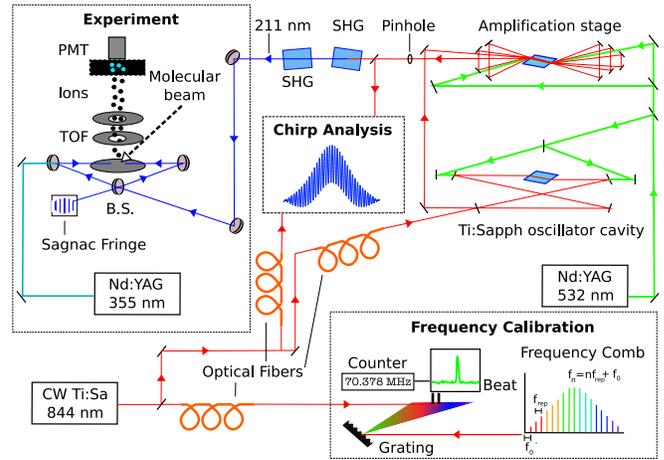}
\caption{(Color online) A schematic layout of the experimental setup. The oscillator cavity is seeded by a cw Ti:Sa laser, the pulsed output of which makes multiple passes in an amplifier stage. The amplified output is frequency up-converted in two frequency doubling (SHG) stages leading to fourth harmonic generation of $\sim$211 nm. The deep UV radiation is sent to the experiment, where molecules in the \xstate\ $v'=1$ state, populated by electrical discharge, are optically excited in a two-photon Doppler-free configuration . The cw-seed light is compared to a frequency comb while the frequency offset between pulsed and cw-seed light is measured via on-line chirp analysis to obtain an absolute frequency calibration. See text for further details.}
\label{Fig:Setup}
\end{figure}

\begin{figure}[b]
\includegraphics[width=1\linewidth]{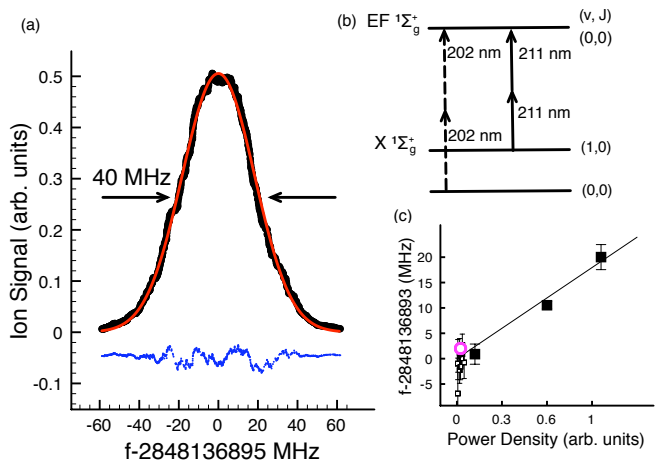}
\caption{(Color online) (a) Recording of the $EF^{1}\Sigma_{g}^{+}-X^{1}\Sigma_{g}^{+}$(0,1) Q(0) two photon transition in H$_{2}$ (in black). Below (in blue) the residuals from a fit to a Gaussian profile (red line) are shown. (b) Partial level scheme of molecular hydrogen showing the measurement principle to obtain the ground tone vibrational splitting as a combination difference. The transitions measured in the present study are indicated by solid arrows while those measured in a separate study \cite{Hannemann2006, Salumbides2008} are shown as dashed arrows. (c) Assessment of the AC-Stark effect in a plot of the absolute frequency of the $EF-X$(0,1) Q(0) transition in H$_{2}$ vs. the 211 nm power density in the interaction zone. The open circle signifies the result shown in (a).}
\label{Fig:Q0}
\end{figure}

The present study focuses on a precise laser spectroscopic measurement of the rotationless fundamental quantum of vibration in H$_2$, HD and D$_2$. In the absence of rotation a one-photon transition between the ($v''=0,J''=0$) and ($v'=1,J'=0$) quantum states is strictly forbidden by nonrelativistic quantum mechanics.  It is the nuclear spin-rotation coupling which makes this transition possible for D$_{2}$ and HD, but the oscillator strength is extremely weak, while it remains zero for H$_{2}$. Therefore an experimental approach is adopted to measure this quantity via the combination difference of \emph{separate} two-photon transitions involving a \emph{common} electronically excited state ($EF^1\Sigma_g^+$, $v=0$, $J=0$) in a setup as displayed in Fig.~\ref{Fig:Setup}. After a previous two-photon study of the Q(0) line in the $EF-X$(0,0) band of the H$_2$, HD and D$_2$ isotopes \cite{Hannemann2006,Salumbides2008}, now the Q(0) two-photon lines in the $EF-X$(0,1) band are subjected to a frequency metrology experiment. The excitation scheme is indicated in Fig.~\ref{Fig:Q0} (b).

The experiment was performed with a long-pulse (20 ns duration) injection-seeded oscillator-amplifier Titanium Sapphire (Ti:Sa) laser system, specifically designed to deliver narrow bandwidth output (as low as 17 MHz) at its fundamental infrared wavelength~\cite{Hannemann2007a}. Its pulsed output was frequency up-converted in two frequency doubling stages to produce the resonant wavelength for inducing the two-photon transitions (211 nm for H$_2$, 209 nm for HD, and 207 nm for D$_2$). The majority of the measurements were performed with a few $\mu$J per pulse of deep UV radiation, while up to 350 $\mu$J per pulse was used to assess the AC Stark effect.
The deep UV radiation is arranged in a Sagnac interferometric configuration of counter-propagating probe laser beams, which facilitates cancellation of the first-order Doppler shift \cite{Hannemann2007}.
The H$_2$ is probed in the collisionless environment of a molecular beam, avoiding the usual difficulties of
collisional frequency shifts in the weak electric quadrupole rovibrational spectra of hydrogen~\cite{Buijs1971, Bragg1982,Cheng2012}. To reduce AC Stark effects a separate 355 nm laser, delayed by 30 ns with respect to the spectroscopy beam, is used to ionize the molecules in a 2+1' resonance-enhanced multi-photon ionization scheme. The molecular ions traverse a time of flight (TOF) tube before being detected. Population of the $X^1\Sigma_g^+, v=1$ state in the molecular beam is achieved through a pulsed pinhole discharge source~\cite{Zhao2011}.

Frequency calibration was carried out via a beatnote measurement of the Ti:Sa seed-frequency against
a fiber-based frequency comb laser, resulting in an uncertainty smaller than 100 kHz. The dominant source of uncertainty is associated with the frequency offset, or chirp, between the cw-seed and pulsed output of the Ti:Sa laser. This phenomenon was previously characterized in detail for the setup~\cite{Hannemann2007a}. For the present measurements the chirp was measured on-line for each laser shot, with average chirp offsets of $-5.00$ MHz with an uncertainty of 0.25 MHz. This leads to an uncertainty of 2 MHz on the $EF-X$ transition frequencies, when a factor of eight is included to take into account the harmonic conversion and two-photon excitation. The systematic uncertainty due to the AC Stark effect was experimentally investigated by performing intensity dependent measurements of the transition frequencies and extrapolating to zero intensity (see Fig.~\ref{Fig:Q0} c).
A listing of the uncertainty budget is given in Table~\ref{Tab:Uncertainty}.

\begin{figure}
\includegraphics[width=1\linewidth]{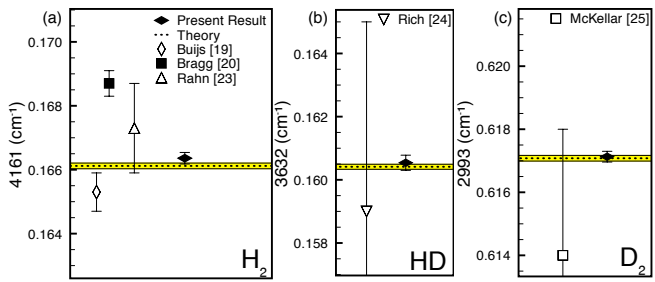}
\caption{(Color online) (a) Current and previously measured values for the fundamental ground tone vibration in H$_{2}$~\cite{Buijs1971,Bragg1982,Rahn1990}. The dashed horizontal line represent the value (and the yellow shaded area represents $\pm 1\sigma$ uncertainties) from the full \textit{ab initio} calculations including QED effects. (b) and (c) represent measured values for the ground tone of HD and D$_{2}$ respectively, along with the full \textit{ab-initio} calculations. These are compared with the results of Rich \textit{et al.} \cite{Rich1982} and McKellar and Oka \cite{McKellar1978} for HD and D$_{2}$ respectively. }
\label{Fig:Splittings}
\end{figure}

A recording of the Q(0) two-photon line for H$_{2}$ is shown in Fig.~\ref{Fig:Q0} (a).
The outcome of the present metrology experiments on the $EF-X$(0,1) transitions in the three hydrogen isotopomers
is given in the lowest section of Table~\ref{Tab:Uncertainty}; combining this with results for the $EF-X$(0,0) band \cite{Hannemann2006,Salumbides2008} then
yields the experimental values for the ground tone frequencies of H$_2$, HD and D$_2$ listed in Table~\ref{Tab:Splittings}. The experimental uncertainties are at the level of $2 \times 10^{-4}$ cm$^{-1}$. This may be compared (see also Fig.~\ref{Fig:Splittings}) with the values extrapolated from direct infrared spectroscopy, yielding reasonable agreement with~\cite{Buijs1971}. However there is a large disagreement at the $6\sigma$-level with respect to the results of Bragg~\emph{et al.}~\cite{Bragg1982}, which were considered the most accurate measurements to date.
Good agreement is found with the less accurate results from Raman spectroscopy \cite{Rahn1990}, while the extrapolations from infrared measurements of HD \cite{Rich1982} and D$_{2}$ \cite{McKellar1978} compare favourably with the present determination.

\begin{table}
\caption{Estimated systematic and statistical uncertainties for the frequency calibrations of the \efstate ($v'=0$) - \xstate ($v''=1$) transitions in \Hm , HD and \Dm\ (in MHz). In the lowest section the Q(0) two-photon transition frequencies are listed in \wn . }
\label{Tab:Uncertainty}
\begin{threeparttable}
\begin{tabular}{l c c}
\\
\hline
\hline
Contribution&Species&Uncertainty (MHz)\\[1mm]
\hline
(i) AC-Stark\tnote{a}&\Hm &0.5\\
 &HD&0.4 \\
 &\Dm & 0.8\\
(ii) DC-Stark& &$<$0.1\\
(iii) Chirp\tnote{a}& &2\\
(iv) Frequency Calibration& & 0.1 \\
(v) Residual Doppler& \Hm & 1.1\\
 &HD& 0.7\\
 &\Dm & 0.6\\[1mm]
 \hline
(vi) Statistics&\Hm &1.5\tnote{b} \\
 &HD&1.6\tnote{c}\\
 &\Dm &2.2\tnote{d}\\
\hline
Combined Uncertainty\tnote{e}& &\\[1mm]
\hline
&\Hm &2.8   \\
&HD&2.7\\
&\Dm &3.2\\
\hline
\hline
Transitions& & (\wn )\\
\hline
$EF-X$(0,1) Q(0)&\Hm &95003.62059(10) \\
 &HD&95669.18608(10)  \\
 &\Dm &96467.83195(10) \\
 $EF-X$(0,0) Q(0)\tnote{f}&\Hm &99164.78691(15) \\
 &HD&99301.34662(23)  \\
 &\Dm &99461.44908(13) \\
 \hline
 \hline
\end{tabular}

\begin{tablenotes}
\footnotesize
\item[a] Chirp and AC Stark offsets are corrected for and not indicated in the table.
\item[b] 1$\sigma$ statistical error based on 63 measurements.
\item[c] 1$\sigma$ statistical error based on 69 measurements.
\item[d] 1$\sigma$ statistical error based on 44 measurements.
\item[e] Quadrature sum of errors (i)-(v).
\item[f] From previous studies~\cite{Hannemann2006,Salumbides2008}.
\end{tablenotes}
\end{threeparttable}
\end{table}

To compare with theory, calculations of molecular hydrogen level energies $E$ are performed in the framework of nonrelativistic quantum electrodynamics (NRQED) making use of an evaluation in orders of the electromagnetic coupling constant $\alpha$
\begin{equation}
E(\alpha) = \,\bigl[{\cal E}^{(0)} + \alpha^2\,{\cal E}^{(2)} + \alpha^3\,{\cal E}^{(3)} + \alpha^4\,{\cal E}^{(4)} \ldots\bigr].
\end{equation}
The nonrelativistic energy ${\cal E}^{(0)}$ is obtained by first producing a Born-Oppenheimer potential
with 15 digit accuracy \cite{Pachucki2010a}, and then solving the Schr\"odinger equation and calculating adiabatic and
nonadiabatic corrections perturbatively in powers of the electron-nucleus mass ratio \cite{Pachucki2009}.
The resulting nonrelativistic binding energies are accurate to a few parts in $10^{-4}$ cm$^{-1}$ \cite{Pachucki2009}
and are in excellent agreement with the direct nonadiabatic calculations (a variational approach) for \textit{rotationless} molecular hydrogenic levels,
performed by Adamowicz and co-workers for the case of H$_{2}$~\cite{Stanke2008}.
For HD and D$_2$ the agreement is less perfect \cite{Stanke2009,Bubin2011}, which we attribute to the need for much larger basis sets used in the direct nonadiabatic calculations (\emph{i.e.} without relying on a Born-Oppenheimer approach), in particular for HD where the breaking of $u-g$ inversion symmetry must be covered in the basis set. Moreover for heavier masses the calculations converge more slowly~\cite{Frolov2003}.
The cancellation of errors for close-lying levels, probing the same part of the potential,
leads to a significant improvement of uncertainties for the \textit{rotationless} levels to $1 \times 10^{-5}$ (see Table~\ref{Tab:Splittings}).

Leading relativistic corrections ${\cal E}^{(2 )}$ are calculated from the expectation value of
the Pauli Hamiltonian \cite{Piszczatowski2009,Komasa2011}.
The leading QED corrections  ${\cal E}^{(3)}$, well-known in the hydrogen and helium atoms \cite{Drake2000},
can be evaluated as expectation values of more complicated operators, such as
the Bethe logarithm \cite{Drake2000} to an accuracy of $10^{-6}$ cm$^{-1}$.
The quoted uncertainties in ${\cal E}^{(3 )}$ originate from neglecting nonadiabatic and relativistic recoil corrections.
The main issue at present is the ${\cal E}^{(4)}$ QED correction, producing the largest contribution to the uncertainty.
For molecular hydrogen this term has not been calculated explicitly due to the high complexity of NRQED operators.
Nevertheless, on the basis of results for the hydrogen and helium atoms, numerically leading contributions
can be represented by calculable Dirac $\delta$-functions. The remaining contribution, dominated by the electron self-energy, is represented by a correction to the Born-Oppenheimer potential, thus contributing to all molecular levels. A conservative uncertainty of 50\% is estimated for contributions related to the energy difference between vibrational levels.   The radial nuclear functions for $v=0$ and $v=1$ probe almost the same range of internuclear distance,
leading to significant cancellation of the uncertainty in ${\cal E}^{(4)}$. This results
in final theoretical predictions at an accuracy $1 \times 10^{-4}$ cm$^{-1}$ for the full QED evaluation
of the rotationless fundamental ground tones in H$_2$, D$_2$ and HD.
All contributions to the energies and their uncertainties are presented in Table~\ref{Tab:Splittings} alongside
the experimental results.

The present experiment finds excellent agreement between the measured values of the
fundamental ground tones of the H$_2$, HD, and D$_2$  molecules and a full \emph{ab initio}
calculation including nonadiabatic, relativistic and quantum electrodynamical effects at a combined
precision level of $2 \times 10^{-4}$ cm$^{-1}$. Implicitly included are calculations of electron correlations, a phenomenon known to pose a major difficulty in quantum \textit{ab initio} calculations of molecular structure.
The comparison is graphically represented in Fig.~\ref{Fig:Splittings}. The present comparison is an improvement over the previous test of QED on the \Hm\ dissociation energy~\cite{Liu2009} (limited by the calculations to $1.2 \times 10^{-3}$ cm$^{-1}$=36 MHz~\cite{Piszczatowski2009}) and over a test of high rotational states in H$_{2}$~\cite{Salumbides2011} (limited by the experimental values at $5 \times 10^{-3}$ cm$^{-1}$=150 MHz). The result constitutes an accurate test of QED in a chemically-bound system.

Although QED is considered to be the best tested theory to date, a recent disagreement on the proton size between muonic hydrogen and atomic hydrogen experiments opens a perspective on additional interactions between leptons and hadrons beyond the realms of QED~\cite{Pohl2010,Antognini2013}. In view of this development, the high level of agreement between the most accurate theory and experiment for the molecular hydrogen level energies may be interpreted to constrain effects of possible long-range hadron-hadron interactions. Theories invoking extra compactified dimensions \cite{Hamed1998}, or predicting new particles and new interactions, of which a long-range interaction between hadrons is a possibility, may be parametrized by a Yukawa-type potential $V(r)=N^{2} \frac{\beta}{r}e^{(-r/\lambda)}$, where $N$ is the nucleon number, $\beta$ is a coupling constant and $\lambda$ is the effective range of the interaction. In the approximation $\lambda \gg R=0.74$ \AA\ the present data yield a constraint on long-range hadron-hadron interactions, quantified by $\beta<6\times10^{-8}$ eV$\cdot$\AA . In this sense precision molecular spectroscopy opens an avenue to search for new physics.

\begin{table}
\squeezetable
\caption{Summary of quantitative results with 1$\sigma$ uncertainties given in brackets ();
all values in cm$^{-1}$.
$\Delta^{\mathrm{Exp}}_{01}$ represent the experimental values for the vibrational ground tones.
The theory values are given for separate contributions of Born-Oppenheimer energies $\Delta^{\mathrm{BO}}_{01}$,
adiabatic corrections $\Delta^{\mathrm{Adiab}}_{01}$ and nonadiabatic corrections $\Delta^{\mathrm{Nonadiab}}_{01}$ leading to a value for nonrelativistic energy $\Delta^{\mathrm{Non-rel}} $. 
Relativistic effects $\Delta^{\mathrm{Rel}}_{01}$ and QED effects (leading $\Delta^{\mathrm{QED}}_{01}$ and higher order $\Delta^{\mathrm{HQED}}_{01}$) are added to form a final theoretical value for the ground tones: $\Delta^{\mathrm{Th}}_{01}$. Note that for each contribution, the cancellation of uncertainties for the level energies of $v=1$ and $v=0$ states is included. The bottom row lists the difference between experiment and theory for the fundamental ground tone vibration in the three hydrogen isotopomers.}
\label{Tab:Splittings}
\begin{center}
\begin{tabular}{l r@{.}l r@{.}l r@{.}l }
\hline
\hline
\multicolumn{7}{c}{\textit{\textbf{Experiment}}}\\
\hline
 &\multicolumn{2}{c}{H$_2$}&\multicolumn{2}{c}{HD}&\multicolumn{2}{c}{D$_2$} \\
\hline
$\Delta^{\mathrm{Exp}}_{01}$&4161&16632(18) &3632&16054(24) & 2993&61713(17) \\
\hline
\hline
\multicolumn{7}{c}{\textit{\textbf{Theory}}}\\
\hline
 &\multicolumn{2}{c}{H$_2$}&\multicolumn{2}{c}{HD}&\multicolumn{2}{c}{D$_2$}\\
 \hline
 \multicolumn{7}{c}{\textit{Non-relativistic contributions}}\\
$\Delta^{\mathrm{BO}}_{01}$&4163&40350&3633&71956 &2994&44084\\
$\Delta^{\mathrm{Adiab}}_{01}$&-1&40284&-0&93259 &-0&52150\\
$\Delta^{\mathrm{Nonadiab}}_{01}$&-0&83649& -0&62872&-0&30447\\
\hline
$\Delta^{\mathrm{Non-rel}}$&4161&16416(1)&3632&15826(1)&2993&61487(1)\\
\hline
\multicolumn{7}{c}{\textit{Relativistic and QED effects}}\\
$\Delta^{\mathrm{Rel}}_{01}$ ($\alpha^{2}$)&0&02341(3)& 0&02093(2)&0&01771(2)\\
$\Delta^{\mathrm{QED}}_{01}$ ($\alpha^{3}$)&-0&02129(2)&-0&01863(2) &-0&01539(2)\\
$\Delta^{\mathrm{HQED}}_{01}$ ($\alpha^{4}$)&-0&00016(8)&-0&00014(7) &-0&00012(6)\\
\hline
$\Delta^{\mathrm{Th}}_{01}$&4161&16612(9)&3632&16041(8)&2993&61708(7)\\
\hline
\hline
\multicolumn{7}{c}{\textit{\textbf{Comparison}}}\\
\hline
 &\multicolumn{2}{c}{H$_2$}&\multicolumn{2}{c}{HD}&\multicolumn{2}{c}{D$_2$}\\
 \hline
 $\Delta^{\mathrm{Exp}}_{01}$-$\Delta^{\mathrm{Th}}_{01}$&0&00020(20)&0&00013(25)&0&00005(18)\\
\hline
\hline
\end{tabular}
\end{center}
\end{table}

This research was supported by the Netherlands Foundation for Fundamental Research of Matter (FOM) through the program "Broken Mirrors \& Drifting Constants". We also acknowledge support by the NCN grants N-N204-015338 (J.K.) and
2012/04/A/ST2/00105 (K.P.)
as well as by a computing grant from Poznan Supercomputing and
Networking Center, and by PL-Grid Infrastructure.


\bibliographystyle{apsrev4-1}
\bibliography{/Users/garydee/Documents/Articles/CompleteDataBase}

\end{document}